\documentclass[aps,amsmath,amssymb,11pt,superscriptaddress,showkeys]{revtex4-2}
\usepackage[english]{babel}
\usepackage[utf8]{inputenc}
\usepackage{comment}
\usepackage{amsmath}
\usepackage{mathtools}
\usepackage{setspace}

\usepackage{amsfonts}
\usepackage{amssymb}
\usepackage{amsmath}
\usepackage{pifont,MnSymbol}
\usepackage{float}
\usepackage{placeins}
\usepackage{graphicx}
\usepackage[normal,normal,bf]{caption}
\usepackage{subfigure}
\usepackage[export]{adjustbox}
\usepackage{color}
\usepackage{tabularx}
\usepackage{multirow}
\usepackage{tabularx}
\usepackage{float}
\usepackage[colorinlistoftodos, color=green!40, prependcaption]{todonotes}
\usepackage{amsthm}
\usepackage{mathtools}
\usepackage{physics}
\usepackage{xcolor}
\usepackage{graphicx}
\usepackage[left=23mm,right=13mm,top=35mm,columnsep=15pt]{geometry} 
\usepackage{adjustbox}
\usepackage{placeins}
\usepackage[T1]{fontenc}
\usepackage{lipsum}
\usepackage{csquotes}
\usepackage[pdftex, pdftitle={Article}, pdfauthor={Author}]{hyperref} 

\DeclareMathAlphabet{\mathbfit}{OML}{cmm}{b}{it}
\newcommand{\n}{\mathbfit{n} }

 \newcommand{\grads}{\mathbf{\nabla}_{\!s} }
 
  \newcommand{\Vs}{\mathbfit{v}_{\!s} }

\newcommand{\Ds}{{{\it D}_{\!s}}}

 \newcommand{\Gammainf}{{\mathrm{\Gamma_{\!\infty}}} }

  \hypersetup{
    bookmarks=true,         
    unicode=false,          
    pdftoolbar=true,        
    pdfmenubar=true,        
    pdffitwindow=false,     
    pdfstartview={FitH},    
    colorlinks=true,       
    linkcolor=blue,          
    citecolor=red,        
    filecolor=magenta,      
    urlcolor=blue           
}

\begin{document}
\title{The presence of surfactants controls the stability of bubble chains in carbonated drinks}

\author{Omer Atasi}
\affiliation{Institut de M\'ecanique des Fluides de Toulouse, Universit\'e de Toulouse, CNRS, Toulouse, France}
\author{Mithun Ravisankar}
\affiliation{Center for Fluid Mechanics, School of Engineering, Brown University, Providence RI 02912, USA}
\author{Dominique Legendre}
\affiliation{Institut de M\'ecanique des Fluides de Toulouse, Universit\'e de Toulouse, CNRS, Toulouse, France}
\author{Roberto Zenit}
    \email[Corresponding author: ]{roberto_zenit@brown.edu (Roberto Zenit)}
    \affiliation{Center for Fluid Mechanics, School of Engineering, Brown University, Providence RI 02912, USA}

\date{\today} 

\begin{abstract}
Bubbles appear when a carbonated drink is poured in a glass. Very stable bubble chains are clearly observed in champagne, showing an almost straight line from microscopic nucleation sites from which they are continuously formed. In some other drinks such as  soda, such chains are not straight (not stable). Considering  pair interactions  for spherical clean bubbles, bubble chains should not be stable which contradicts these observations. The aim of this work is to explain the conditions for bubble chain stability. For this purpose, experiments and direct numerical simulation are conducted. The bubble size as well as the level of interface contamination are varied, to match the range of parameters in typical drinks. Both factors are shown to affect the bubble chain stability. The transition from stable to unstable behavior results from the reversal of the lift force, which is induced by the bubble wake. A criteria based on the production of vorticity at the bubble surface is proposed to identify the conditions of transition from stable to unstable bubble chains. Beyond carbonated drinks, understanding bubble clustering has impact in many two-phase problems of current importance.
\end{abstract}

\maketitle

\section{Introduction}
Flavour and smell are not the only sensorial traits of drinking champagne.  The bubble chains that appear in a tall glass complete the trifecta: the beauty and visual satisfaction of observing the consecutive and ordered formation of bubbles are an essential part of the experience. Bubbles in carbonated drinks, not just champagne, form as excess Carbon Dioxide exsolves from the supersaturated liquid. The supersaturation condition is easily achieved since the bottling pressure is higher to atmospheric, hence more gas can be dissolved in the liquid. The chains form as bubbles are continuously formed at a certain nucleation site; once the bubble reaches a certain critical size for which buoyancy overcomes the surface tension force that keeps the bubble attached, the bubble begins to ascend freely. Immediately after the bubble has been dislodged, a new one being to form. This process continues for several minutes until most of the dissolved gas comes out of solution \cite{liger-belair_physics_2005}. The speed at which bubble rise in the liquid depends on their size and the liquid properties \cite{tripathi_dynamics_2015,hayashi_lift_2020,senapati_numerical_2019}. Once the bubbles reach the liquid surface they briefly remain at it while the liquid film drains \cite{aradian_marginal_2001,lhuissier_bursting_2012,champougny_life_2016,atasi_lifetime_2020, rage_bubbles_2020}. When the liquid film becomes sufficiently thin, it spontaneously bursts with a characteristic sound \cite{Poujol2021} and forming aerosol droplets \cite{rage_bubbles_2020,liger-belair_unraveling_2009,ghabache_evaporation_2016}. All these processes make a glass of carbonated liquid a complex laboratory of two-phase flows with intricate physico-chemical processes \cite{liger-belair_physics_2005,ghabache_evaporation_2016,liger-belair_recent_2008,liger-belair_unraveling_2009, zenit_fluid_2018}.

The present investigation focuses on the stability of bubble chains:  as the bubbles form consecutively from the same nucleation site, should they remain in-line or not?  Stable in-line motion is commonly observed in Champagne and other carbonated wines, but is this what one should expect by considering the hydrodynamic interactions among bubbles? The flow disturbance created by the motion of a single bubble ascending  will exert certain influence on a subsequent bubble moving behind.  Beyond the curiosity of bubble chains in drinks, these results are important in clarifying the key mechanisms that lead to bubble distributions in bubbly flows in general. The nature of hydrodynamic interactions among bubbles determines whether or not cluster are formed \cite{Velez2014} which, in turn, greatly influence the mixing and agitation levels of bubbly flows\cite{risso2018agitation}.

Let us first consider the case of a two spherical bubbles, with clean interfaces, rising in-line. If the Reynolds number (defined in the Methods section), is higher than 10,  the in-line motion is unstable because bubbles are ejected from the wake of any previous bubble. The process is explained by the wake of the leading bubble. Its interaction with the trailing bubble results in the appearance of a lift force \cite{legendre_lift_1998, hallez_interaction_2011}, $F_L$: 
\begin{equation}
\mathbfit{F}_L= C_L \rho_l \frac{4\pi R^3}{3} \left(\mathbf{V}_L - \mathbf{V}_B \right) \times \left( \nabla \times  \mathbf{V}_L \right)
\label{eqn:lift}
\end{equation}
where $\mathbf{V}_L$ is the velocity field of the leading bubble wake, $\mathbf{V}_B $ is the velocity field of the trailing bubble and $C_L$ is a lift coefficient (positive) for spherical and clean bubbles. Such unstable in-line bubble interaction has been confirmed by experiments \cite{Kusuno2019} and simulations \cite{hallez_interaction_2011}. Recent simulations have also revealed that the interaction between two bubbles, in the inertial regime, is full of subtleties \cite{zhang2021,zhang2022}.

For the case of bubbles in champagne, and other aqueous carbonated drinks, the bubble Reynolds number increases monotonically after the bubble formation as the bubble continues to grow. However, and most importantly, it reaches a value of $Re>10$ a few millimeters after formation \cite{Liger-Belair2000}. Hence, one would expect that stable bubble chains would not be observed in a glass champagne. In this investigation we argue that a stable bubble chain is possible only if the lift force acting on the trailing bubble changes sign. In such a case, the bubbles would remain in-line regardless of small disturbances.

Previous investigations have addressed the lift reversal process. Two mechanisms are known to cause lift reversal, bubble deformation  \cite{Tomiyama_2002, hayashi_lift_2020, hayashi_lift_2021} and surface contamination \cite{takagi2008}. Either one, or both, can lead to a reversal of the lift force. Harper \cite{Harper2008} envisioned that surfactants were in fact responsible for the stabilization of bubble chains observed on common beverages, but did not propose a physical mechanism to explain the stabilization.

The main objective of this work is to clarify the origin of the in-line stabilisation of bubble chain, inspired by the particular case of bubble chains in carbonated drinks. Experiments and numerical simulations are conducted to vary the bubble size and surface contamination independently. To simplify the system, we consider non-growing bubbles. In the case of Carbon Dioxide bubbles rising in supersaturated liquids, the bubbles grow as they ascend. Since the grow rate, $\dot R$, is smaller than the bubble velocity $\dot R/U_B<1$, the added mass effect can be neglected \cite{Liger-Belair2000}.

\section{Materials and Methods}

\subsection{Relevant dimensionless groups}

The problem of a bubble of radius $R$ rising steadily in a fluid of viscosity $\mu$ and density $\rho$ under the action of gravity $g$, with a gas-liquid interfacial tension $\sigma_o$, is completely characterized by two non dimensional numbers. We consider here the Bond number and Archimedes numbers:
\begin{align}
    Bo=\frac{\rho g R^2}{\sigma_0}, ~Ar=\frac{\rho^2g R^3}{\mu^2}.
\end{align}
Note that the Archimedes number is in fact a Reynolds number 
\begin{align}
    Re=\frac{\rho R U_B}{\mu}
\end{align}
where characteristic velocity $U_B=\rho g R^2/\mu$ is obtained from the balance of the bubble viscous drag with buoyancy. The bubble deformation can be quantified considering the bubble aspect ratio  defined as $\chi=D_L/D_S$ where $D_L$ and $D_S$ are the long and short diameters of an ellipsoidal bubble. It is well known that $\chi$ is a function of $Bo$ \cite{legendre2012deformation}.

The transfer of surfactants is characterized by the bulk and the interface P\'eclet numbers, respectively defined as:
\begin{align}
Pe=\frac{\rho g R^3}{\mu D}, ~Pe_s=\frac{\rho g R^3}{\mu D_s}, 
\end{align}
where $D$ and $D_s$ are the surfactant difusivity coefficient in the bulk and the bubble surface, respectively. An estimation of $Pe_s$ for an air bubble of millimetric radius inside a liquid containing fatty acids or other long chain surfactants ($D_s \sim O(10^{-10})$) gives $Pe_s\approx 10^{8}$, indicating negligible surface diffusion. Therefore, $D_s$ has been set to zero in the simulations. 

Furthermore, the surfactant presence at the interface  introduces three additional non dimensional numbers 
\begin{align}
\alpha=\frac{k_a C_0 \mu}{\rho g R}, ~La=\frac{k_a C_0 }{k_d} \text{~and~} M\!a=\frac{\mathcal{R}T \Gamma_\infty}{\sigma_0},
\end{align}
namely the solubility parameter $\alpha$, the Langmuir number $La$ and the Marangoni number $Ma$. $C_0$ is the surfactant
concentration in the liquid bulk, $\Gamma_\infty$ is the maximum concentration of
surfactants at the interface and $k_a$ and $k_d$ are the adsorption and desorption kinetic constants, respectively. $\mathcal{R}$ is the ideal gas constant and $T$ is the absolute temperature.
 The Langmuir number $La$ compares the adsoption and desorption source terms at the interface. Clean interfaces are observed for $La \rightarrow  0 $ while  saturated interfaces for $La \rightarrow  \infty $.
 
The level of rigidity of the interface due to the Marangoni effect can be quantified by comparing the {two terms in the right hand side of Eq. \ref{Eq_cont2}, namely} the 
surface tension gradient {$\nabla_I \sigma \sim La Ma  \sigma_0 /R$} and the liquid viscous shear $\tau \sim \mu_L U_B /R$ experienced by a non contaminated bubble. Therefore the `rigidity' parameter $\Lambda$ \cite{atasi_lifetime_2020, champougny_2015} is  
\begin{equation}
\Lambda = \frac{Ma}{Bo}.
\end{equation}
The value of $ La \Lambda$ increases with interface contamination by surfactants, which in turn result in the increase of the interfacial vorticity {as shown in Eq. \ref{Eq_cont23}}.

\subsection{Range of parameters for bubbly drinks}

The density, surface tension and viscosities for water (Pellegrino sparkling water), a lager beer (Tecate, 4.5\% ethanol), sparkling wine (Segura Viudas Brut, 12.5\% ethanol)  and champagne (Charles de Cazanove, 12\% ethanol) were measured using a pycnometer (25ml), a tensiometer (BPT Mobile, KRUSS Scientific Instruments) and rheometer (ARES-G2 Rheometer, TA instruments, concentric cylinders), respectively. The values of these properties are shown in Table \ref{tab:table1}. These test beverages were degassed before testing, by keeping them in a light vacuum environment (1 mbar) for one hour.
\begin{table}[h]
\centering
\caption{\label{tab:table1}
Values of physical properties water (carbonated water, Pellegrino), beer (lager, Tecate) and champagne (sparkling wine, Brut Segura Viudas). The values of the dimensionless numbers were obtained considering bubbles sizes ranging $0.25 < R < 1$ mm.} 
\begin{tabular}{ccccccccc}
Fluid  & $\rho$, kg/m$^3$ & $\mu$, mPa s & $\sigma$, mN/m &  $Bo$  & $Ar $ & $La$ & $Ma$  \\
\hline\hline
Water &   998 & 1.00  & 72.0 & 0.008 - 0.13 & 152 - 9726  & 0 - 0.1 & 0.1\\
Beer &   1124 & 1.36 & 55.0 & 0.013 - 0.20 & 105 - 6700 & 15 - 38 & 0.0014 \\
Sparkling wine &   1100 & 1.47 & 48.0 & 0.014 - 0.23 & 86 - 5493 & 0.2 - 0.7& 0.3\\
Champagne &   992 & 1.44 & 58.0 & 0.010 - 0.17 & 73 - 4653 & 0.2 - 0.7& 0.3\\
\end{tabular}

\end{table}

Considering bubble radii ranging from 0.25 to 1 mm, the expected values of $Ar$,  $Bo$ and $La$ for beer, champagne and sparkling water, are summarized in Table \ref{tab:table1}. 



{To estimate the values of $La$ and $Ma$ for these drinks we considered the type of surfactants and their expected concentrations \cite{Chang1995,Guzman1986}. 
The equilibrium adsorption parameters for the Langmuir isotherm were determined for various surfactants from equilibrium surface tension measurements.}

{For champagne, the surfactants are mainly fatty acids that contribute to the organoleptic sensations \cite{liger-belair_unraveling_2009}. For fatty acids, the Langmuir adsorption isotherm was fitted to the equilibrium surface tension measurements, which resulted in $0.02 \ \text{m}^3/\text{mol}<k_a/k_d<0.07 \ \text{m}^3/\text{mol}$ and $\Gamma_\infty$ = 10$^{-5}$~ \text{mol/m}$^2$ \cite{Chang1995}. Assuming that $C_0=10~\text{mol/m}^3$, $La$ ranges from 0.2 to 0.7 and $Ma \approx 0.3$. Therefore, we estimated that for bubble diameters ranging from 0.25 to 1 mm moving in champagne, the product of the `rigidity parameter' with the Langmuir number is in the range $0.2<La\Lambda<14$.}

{For beer most surfactants are proteins of molecular weight ranging from 5 to 100 kDa \cite{blasco2011proteins,2009SPoB}. The adsorption of proteins to air/liquid interfaces is significantly different than the adsorption of small molecules. Most importantly, this adsorption cannot be described by the Langmuir isotherm. 
Indeed, proteins adsorption requires higher activation energy than fatty acids, once adsorbed protein unfold from their three dimensional structure and create multilayers of adsorbed proteins conferring elastic and viscous properties to the surface. Also, the adsorption of proteins can be considered as irreversible \cite{Guzman1986,makievski_adsorption_1998,li_protein_2021}. 
Nevertheless, attempts to fit adsorption isotherms of proteins with the Langmuir isotherm exists. Surface coverage measurements for albumine (molecular weight of 67 kDa) is one example. These measurements gave $k_a/k_d=1.3 \times 10^4 \ \text{m}^3/\text{mol}$ and $\Gamma_\infty=4\times 10^{-8}  \ \text{mol/m}^2$ \cite{wertz_adsorption_1999,latour_langmuir_2015}. Assuming a concentration of surface active proteins ranging from 0.08 to 0.2 g/l \cite{Chengtuo} and an average molecular weight of $67$ kDa we estimate $0.0012<C_0<0.003 \ \text{mol/m}^3$. Therefore, we obtain $Ma \approx 0.0014$ and $15<La<40$. For bubble from 0.25 to 1 mm in radius, we infer $La \Lambda = Ma \ La/Bo \approx 0.1-1.6$. 
Note that other adsorption isotherms have been derived to describe the adsorption of proteins on air/liquid or liquid/solid interfaces; however, the use and implementation of these isotherms is beyond the scope of this study.
}

Lastly, for water, we arbitrarily assigned small values of $Ma$ and $La$ to denote that a small amount of surfactants are present due to ordinary laboratory conditions.




\subsection{Experimental setup and materials}

Bubble chain experiments were conducted in a rectangular plexiglass tank of $50\times50\times400$ mm$^3$. At the bottom of the tank, a needle was inserted through a self-healing rubber stopper. Blunt edge needles of three different inner diameters, $D_c$, were used to produce bubbles of different sizes ($0.16$ mm, 0.26 mm and 1.55mm). Air was introduced to the needle to produce bubbles using a syringe pump (Harvard Apparatus PHD 2000 series), considering flow rates ranging from 1 to 10 ml/min. The different flow rates resulted in experiments with different bubble formation frequencies which, in turn, resulted in changes in the inter-bubble spacing.

To match the Archimedes and Bond numbers for bubbles ascending in typical beverages,  a mixture of 35-65 \% (by weight) of water-glycerin was used. The resulting fluid had a viscosity of 0.01 Pa s, which was measured using a  rheometer (ARES-G2 Rheometer, TA instruments, concentric cylinders). The properties of the bubbles produced for each capillary are shown in Table \ref{tab:table_props}.
\begin{table}[h]
\centering
\caption{\label{tab:table_props}
Experimental conditions. We used four capillaries with inner diameter $D_c$ for produce bubbles of different radius $R$. }
\begin{tabular}{l|cccc}
$D_c$, mm             & $R$, mm   & $Bo $         & $Ar$     & $La$         \\ \hline\hline
\multirow{3}{*}{0.16, {\color{red} $\circ, \bullet$}  } & 1.1 - 1.5  & 0.18 - 0.39 & 155 - 476 & 0.0 \\
                      & 0.8 - 1.0  & 0.12 - 0.19 & ~~57 - 115 & 0.5  \\
                      & 0.8 - 1.0  & 0.14 - 0.20 & ~~69 - 113 & 1.5  \\ \hline
\multirow{3}{*}{0.26, {\color{blue} $\circ, \bullet$}} & 1.3 - 1.5 & 0.26 - 0.39 & 262 - 474 & 0.0  \\
                      & 1.2 - 1.3  & 0.30 - 0.35 & 237 - 292 & 0.5 \\
                      & 1.3 - 1.4  & 0.38 - 0.44 & 293 - 359 & 1.5 \\ \hline
\multirow{3}{*}{1.55, {\color{black} $\circ, \bullet$} } & 2.0 - 2.1  & 0.63 - 0.74 & ~~993 - 1243 & 0.0 \\
                      & 1.8 - 2.2  & 0.69 - 0.94 & ~~815 - 1310 & 0.5\\
                      & 1.6 - 1.9  & 0.56 - 0.82 & 517 - 908 & 1.5\\
                      \hline
0.05, Spark. Water, {\color{blue} $\filledstar$} & 0.25 & 0.008 & 152 & 0.1 \\
0.05, Spark. Water, {\color{blue} $\filledstar$} & 0.64 & 0.056 & 2610 & 0.1 \\
0.06, Spark. Water, {\color{blue} $\filledstar$} & 0.76 & 0.078 & 4333 & 0.1 \\
0.16, Spark. Water, {\color{blue} $\filledstar$} & 0.98 & 0.128 & 9042 & 0.1 \\
\hline
0.05, Beer, {\color{black} $\smallstar$} & 0.21 & 0.009 & 62 & 27.5 \\
0.05, Beer, {\color{black} $\smallstar$} & 0.38 & 0.028 & 354 & 27.5 \\
0.06, Beer, {\color{black} $\filledstar$} & 0.67 & 0.089 & 2015 & 27.5 \\
0.16, Beer, {\color{black} $\filledstar$} & 0.84 & 0.143 & 4042 & 27.5 \\
\hline
0.05, Spark. Wine, {\color{red} $\smallstar$} & 0.19 & 0.008 & 37 & 0.45 \\
0.06, Spark. Wine, {\color{red} $\smallstar$} & 0.59 & 0.078 & 1128 & 0.45 \\
0.16, Spark. Wine, {\color{red} $\filledstar$} & 0.77 & 0.133 & 2507 & 0.45 \\
\hline
0.05, Champagne, {\color{magenta} $\smallstar$} & 0.39 & 0.026 & 276 & 0.45 \\
0.05, Champagne, {\color{magenta} $\smallstar$} & 0.52 & 0.045 & 654 & 0.45 \\
0.06, Champagne, {\color{magenta} $\filledstar$} & 0.86 & 0.125 & 2985 & 0.45 \\
0.16, Champagne, {\color{magenta} $\filledstar$} & 0.91 & 0.139 & 3506 & 0.45 \\
\hline
\end{tabular}

\end{table}

To calculate the value of $La$, we considered that for SDS aqueous solutions, $k_a/k_d=0$.11 m$^3$/mol, according to \cite{Chang1995}. The critical micellar concentration (CMC) is 10 mM, approximately; for higher concentrations of SDS, we can consider that the surface concentration no longer increases.  Therefore, we assume that $C_\infty=$ 10 mM. Therefore, different amount of SDS were added to the fluid to reach values of $La$ relevant to those expected for bubbly drinks.

A series of experiments were conducted with actual sparkling water, beer, sparkling wine and real champagne, considering bubble sizes similar to those observed in carbonated liquids. We used a glass capillary that was heated and stretched to have a very small inner diameter where the bubbles were formed. The properties of these liquids appear in Table \ref{tab:table1} and the test conditions are in Table \ref{tab:table_props}.

\subsection{Numerical Simulations}

Direct numerical simulations were conducted with the in house JADIM code by solving the Navier-Stokes equations
coupled with the Level-Set method. We refer the reader to  \cite{abadie_combined_2015,atasi_lifetime_2020} for a detailed
description of the method and its validation. Briefly, Navier-Stokes equations are solved for two Newtonian and
incompressible fluids using the finite volume method (second-order accurate in time and space). Continuity
is ensured through a projection method, and the capillary contribution is considered through the classical
Continuum Surface Force method \cite{Brackbill1992}. The interface position is tracked using the Level-Set method where the transport
of the signed distance to the interface is controlled through the re-distancing method. The important aspect
of the considered numerical approach used here is its ability to account for the interface contamination by the presence of surfactant in the liquid bulk. The concentration $C$ of surfactant in the liquid bulk is given by the classical advection-diffusion equation
\begin{equation}\label{Eq:bulk}
\frac{\partial C}{\partial t}+ \mathbfit{v_c} \cdot \grad C= \mathbf{\nabla} \cdot \left(D_c\grad C\right),
\end{equation}
while the surfactant concentration $\Gamma$ at the interface is given by  \cite{stone_simple_1990,cuenot_effects_1997}
\begin{equation}\label{Eq:surface}
\frac{\partial \Gamma}{\partial t}+ \grads \cdot \left( \Vs \Gamma\right)=\Ds \nabla^2_{\!s}   \Gamma+S_{ \Gamma},
\end{equation}
 $D_c$ and $\Ds$ are the diffusion coefficients of the surfactants in the
liquid phase and along the interface, respectively, $\mathbfit{v_c}$ is the velocity field in the liquid phase, $\Vs$ is the projection of $\mathbfit{v_c}$ on
the tangent to the interface,$\grads=\left(\mathbf{ I}-\left(\n \otimes \n\right)\right)\cdot\grad $ is the surface gradient operator  and $S_{\Gamma}$ is  the adsorption/desorption source term of  surfactants at the interface \cite{cuenot_effects_1997,muradoglu_front-tracking_2008},
\begin{equation} \label{Eq:SG}
S_{\Gamma}=k_a C_I \left(\Gammainf- \Gamma\right)-k_d  \Gamma,
\end{equation}
where $k_a$ and $k_d$ are adsorption and desorption kinetic constants, respectively, $\Gammainf$ is the maximum concentration of surfactants on the interfaces, and $C_I$ is the surfactant concentration at the vicinity of the gas-liquid interfaces. {Considering Eqs. \ref{Eq:surface}-\ref{Eq:SG}, in a steady state ($S_{\Gamma}=0$), the bubble interface is covered by the uniform concentration 
\begin{equation} \label{Eq:Gamma0}
\Gamma_0= \Gammainf \frac{La}{1+La}
\end{equation}
This concentration is the relevant concentration for the interface concentration normalization (see Eq.  \ref{Eq_cont23}).}

The variation of the surface tension $\sigma$ with $\Gamma$  is described using the Langmuir adsorption isotherm \cite{Levich1962}:
\begin{equation} \label{Eq:Langmuir}
\sigma = \text{max}\left[\sigma_\infty , \sigma_0 \left(1+ \frac{\mathcal{R} T \Gammainf}{\sigma_0} \ln{\left( 1 - \frac{\Gamma}{ \Gammainf}  \right)} \right) \right]
\end{equation}
$\mathcal{R}$ is the ideal gas constant and $T$ is the absolute temperature. $\sigma_\infty$ is {a threshold surface tension, set to $0.05\sigma_0$ for stability purposes. It has been verified that the exact value of this threshold does not affect the bubbles velocity or trajectory and that the surface tension never reaches this threshold for the range of parameter investigated in this study.}

\section{Results}

\subsection{Experiments}

To characterize the system we consider millimetric-sized Nitrogen bubbles forming in a capillary tube and ascending in different viscous liquids. We vary the Bond, $Bo$, the Archimedes, $Ar$ and the Langmuir, $La$, numbers, defined in the Methods Section. 

\begin{figure}[ht]
\centering
\includegraphics[width=0.7\linewidth]{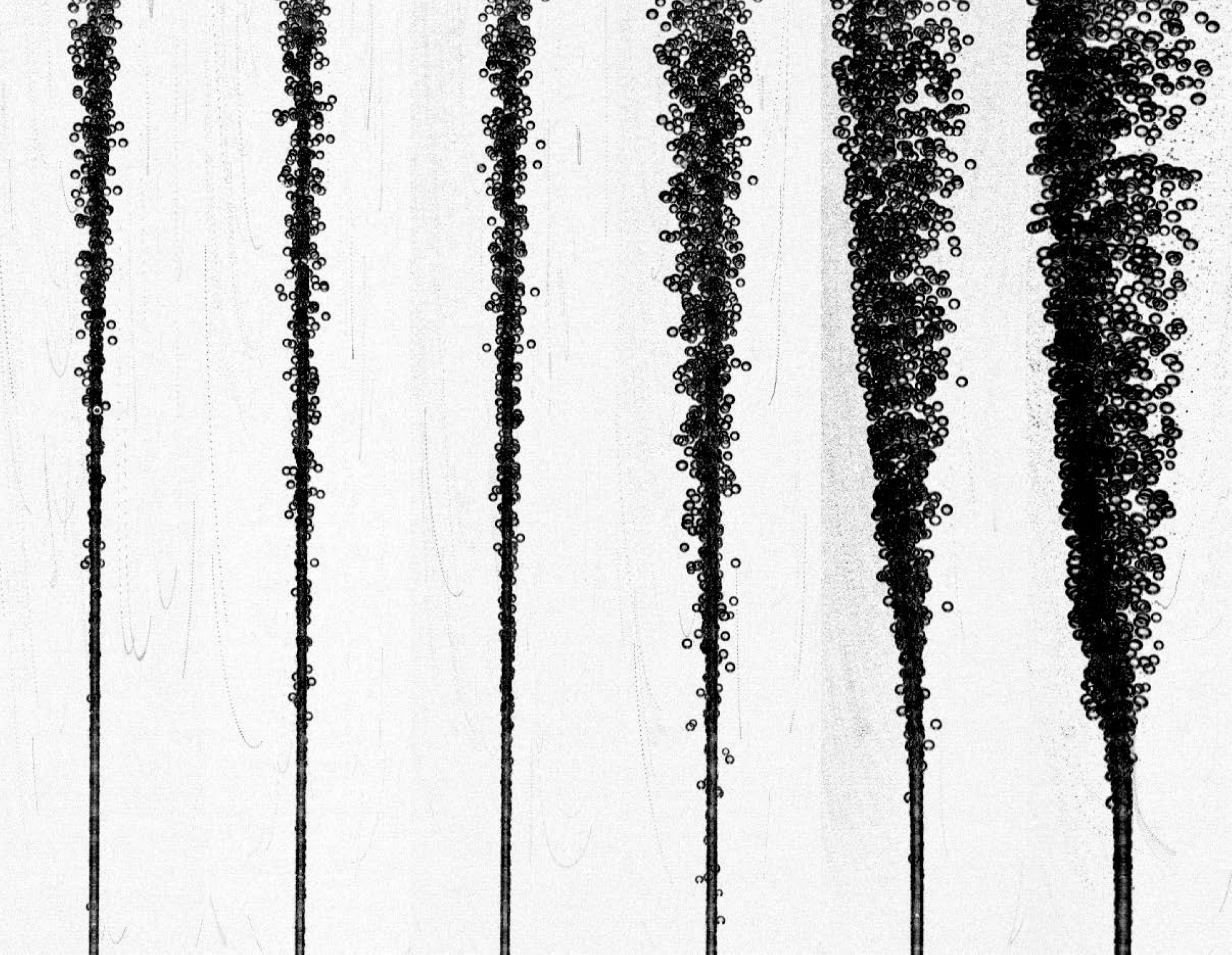}
\caption{Bubble chains forming from a capillary in a clean fluid. The flow rate through the capillary increases from 2 to 10 ml/min, from left to right. The bubble frequency ranges from 6 to 18 bubbles/s. $Ar\approx 150$, $Bo\approx 0.2$ and $La=0$. See videos of these experiments in the Supplemental Material.}
\label{fig:frequency}
\end{figure}

Fig. \ref{fig:frequency} shows images of bubble chains (superposing many different instants) using a capillary tube and a clean water-glycerin mixture, considering different gas flow rates. From similitude arguments, the bubble size, liquid properties and gas flow rate are equivalent to bubbles moving in champagne \cite{Liger1999} and other carbonated beverages\cite{liger2019carbon}. In other words, 
the Archimedes and Bond numbers are matched as closely as possible for this experiments to those of carbonated drinks. The lateral dispersion of bubbles from the point of formation indicates that these chains are not stable. As the gas flow rate increases, the bubble frequency increases and, in turn, the separation distance in between bubbles decreases. The side-ways motion of the bubbles appears sooner as the bubble separation decreases. This is expected, since the hydrodynamic interaction is stronger for shorter distances. 

\begin{figure}[ht]
\centering
\subfigure[$R$ increases, $La\approx 0$]{ \includegraphics[width=0.3\columnwidth]{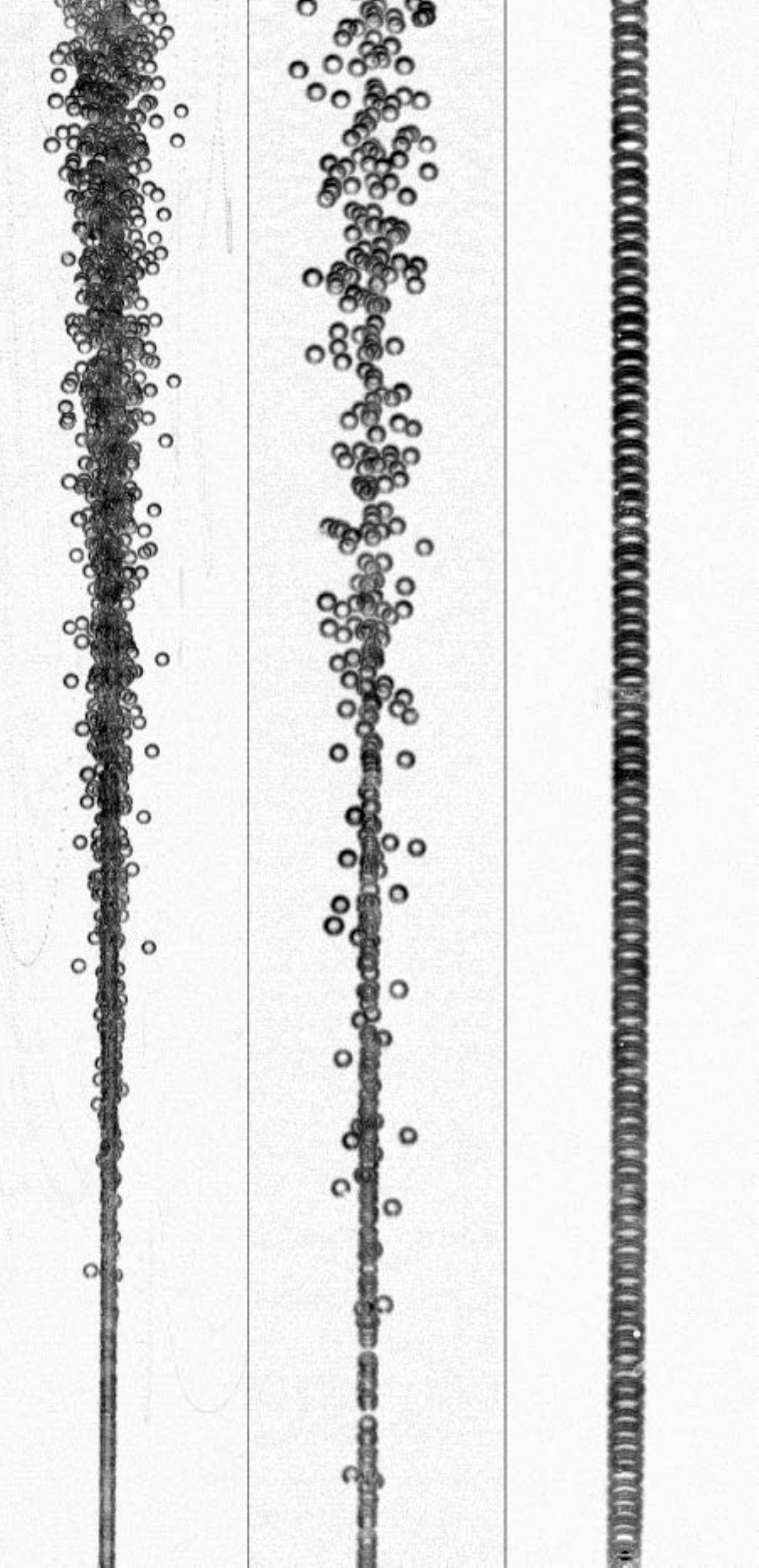}
}
\subfigure[$La$ increases, $R\approx$ constant]{ \includegraphics[width=0.3\columnwidth]{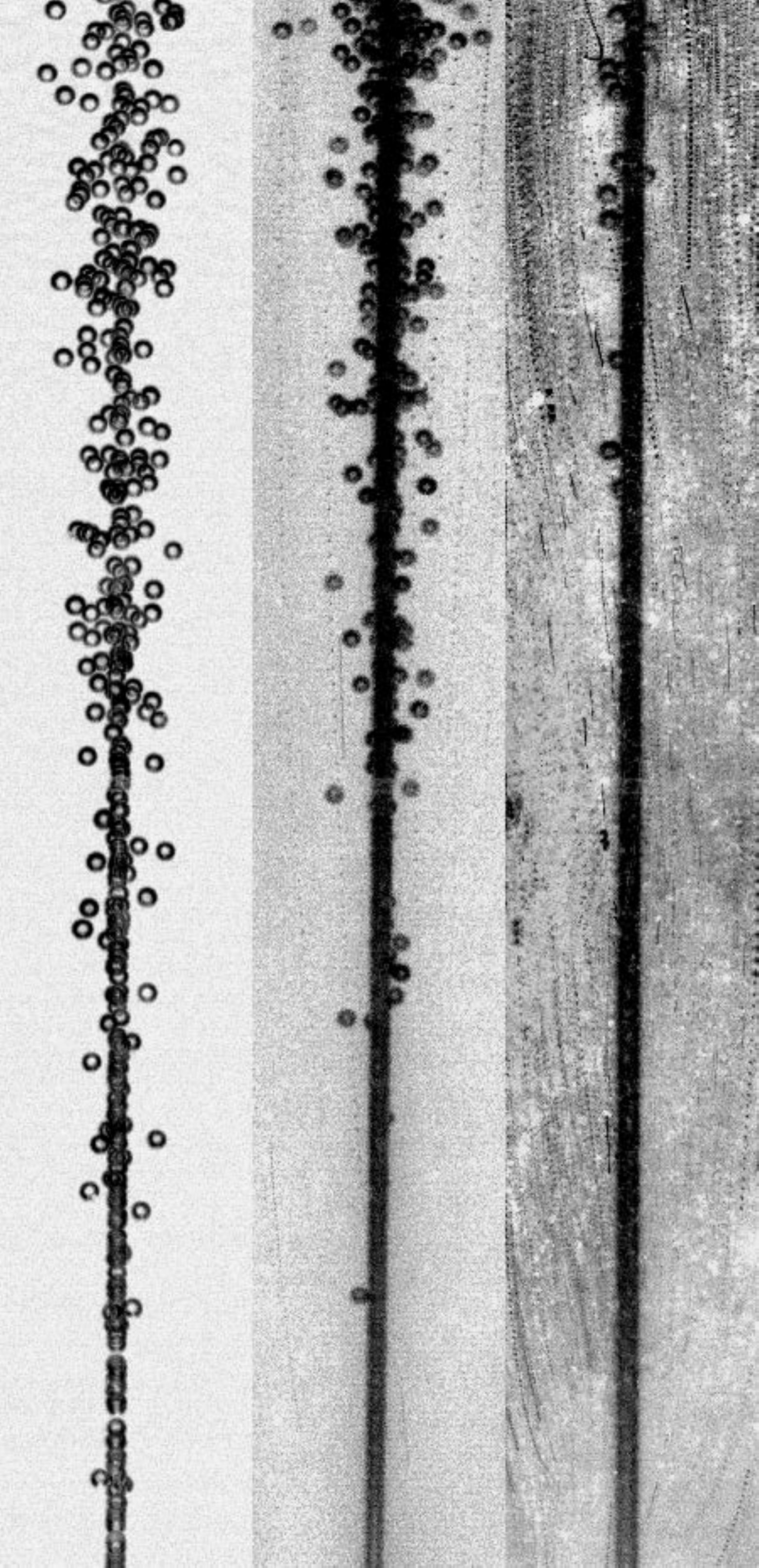}
}
\caption{The effect of bubble size and surface contamination on bubble chain. (a) The bubble size increases (from left to right), for a liquid without surfactants $La=0$; $Ar$=(150, 240, 1050) and $Bo$=(0.18, 0.25, 0.66) for (left, center, right).  (b) The bubble size is approximately the same, but the amount of surfactant increases (from left to right). $La$=(0, 0.1, 0.5), $Ar$=(240, 150, 225) and $Bo$=(0.24, 0.21, 0.32) for (left, center, right). $R$ is the bubble radius. See videos of these experiments in the Supplemental Material.}
\label{fig:exp1} 
\end{figure}

To investigate the reason behind the observed unstable behavior of bubble chains we conducted two types of experiments. First, the bubble size was increased considering the same clean fluid by changing the size of the capillary through which the bubbles are produced. These results are shown in Fig. \ref{fig:exp1} (a). In this case, the bubble chain is unstable for small sizes  (left and middle images) but transitions to a stable in-line chain behavior for large bubbles (right image). Note that, the Bond and Archimedes numbers for these large bubbles are larger than that expected for bubble-chains on champagne and other beverages (see methods section).

In the second experiment, we kept the bubble size relatively fixed by using the same capillary tube; hence, $Bo$ and $Ar$ did not change significantly. But in this case, we gradually added a surfactant (Sodium Dodecyl Sulfate) to increase the surface contamination. Fig. \ref{fig:exp1} (b) shows that as the amount of surfactants increases, the bubble chains transitions from being unstable to stable. Experiments for only one flow rate are shown for both cases (increasing size and increasing surfactant); experiments for other flow rates (not shown) display the same behavior. 

These two experiments clearly indicate that there are two possibilities to stabilize a bubble chain. However, the physical mechanisms that leads to stabilization remains unexplained. To help understand this issue, we conducted numerical simulations. 

\subsection{Numerical Simulations}
The details of the numerical scheme can be found in the Methods section and in the references cited therein. Most importantly, these simulations are capable of modeling the motion of deformable bubbles considering the presence of surfactants. 

The simulations considered a pair of gas bubbles, initially at rest, rising due to buoyancy contained in stagnant fluid domain. To simulate the conditions found in a bubble chain, periodic boundary conditions were imposed on the top and bottom walls of the domain. Many simulations were conducted considering different bubble sizes and contamination levels. The bubble pair is initially placed slightly off the in-line configuration (less the 0.2 $R$).

\begin{figure*}[ht]
\centering
\subfigure[]{ \includegraphics[width=0.31\columnwidth]{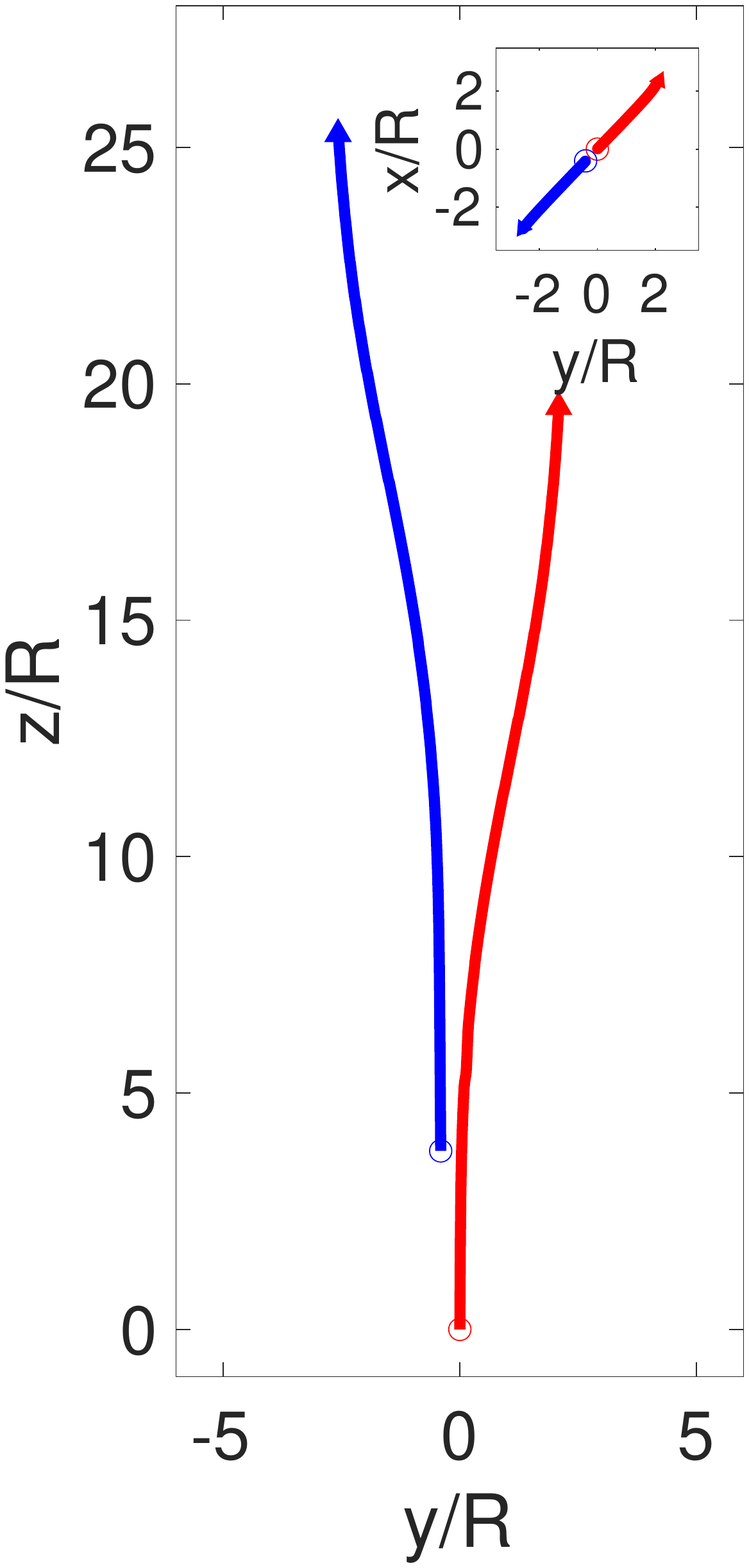}
}
\subfigure[]{ \includegraphics[width=0.15\columnwidth]{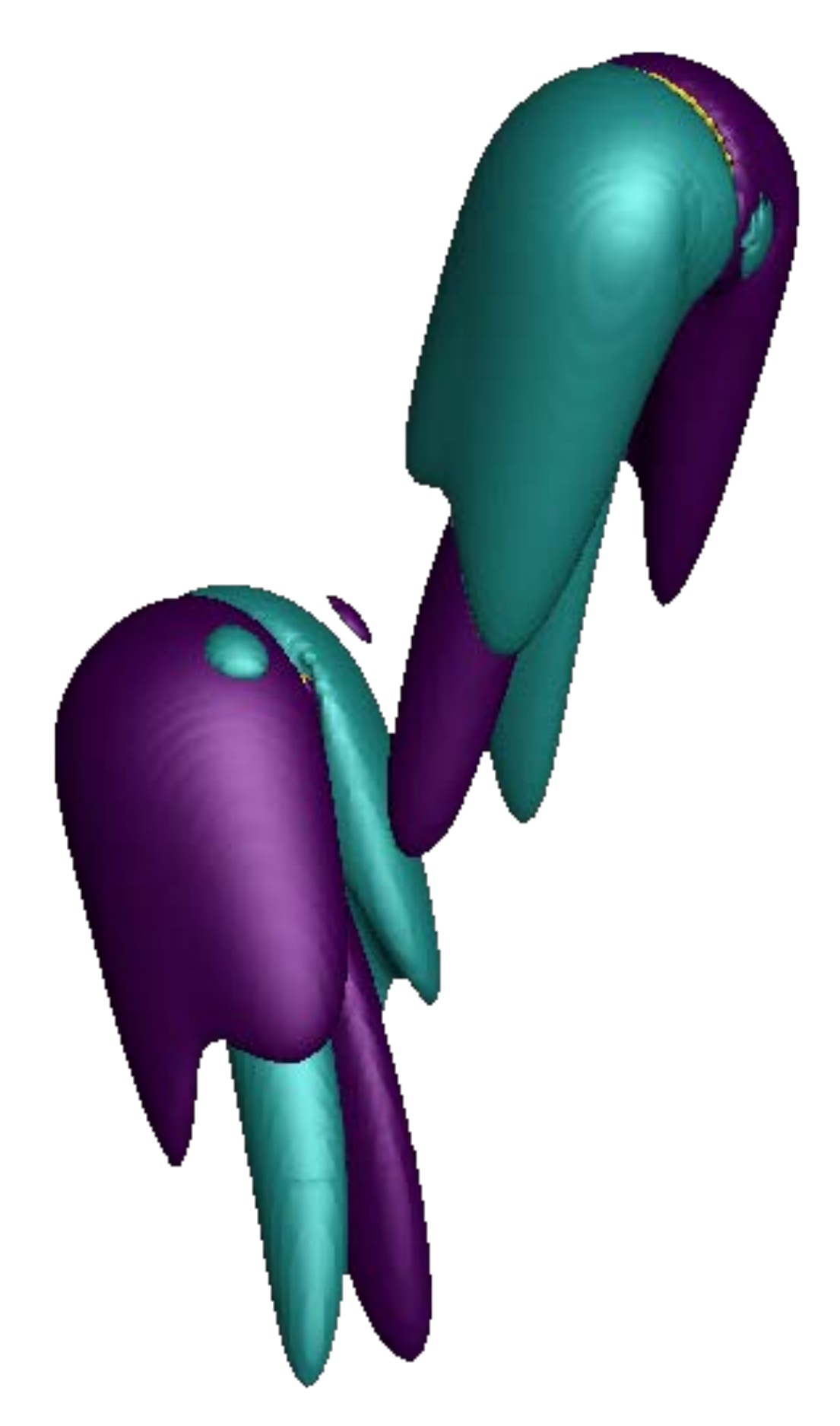}
}
\subfigure[]{ \includegraphics[width=0.31\columnwidth]{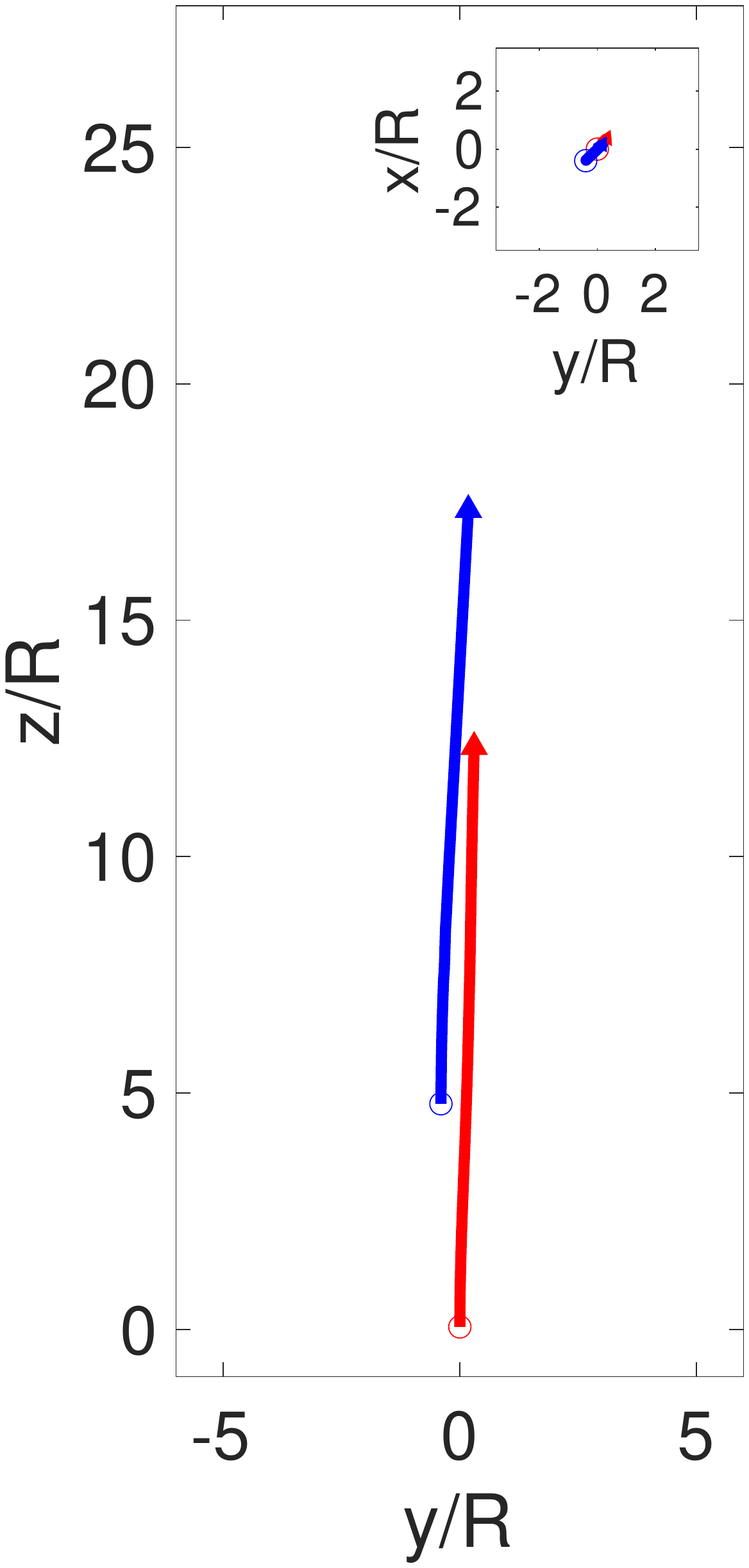}
}
\subfigure[]{ \includegraphics[width=0.15\columnwidth]{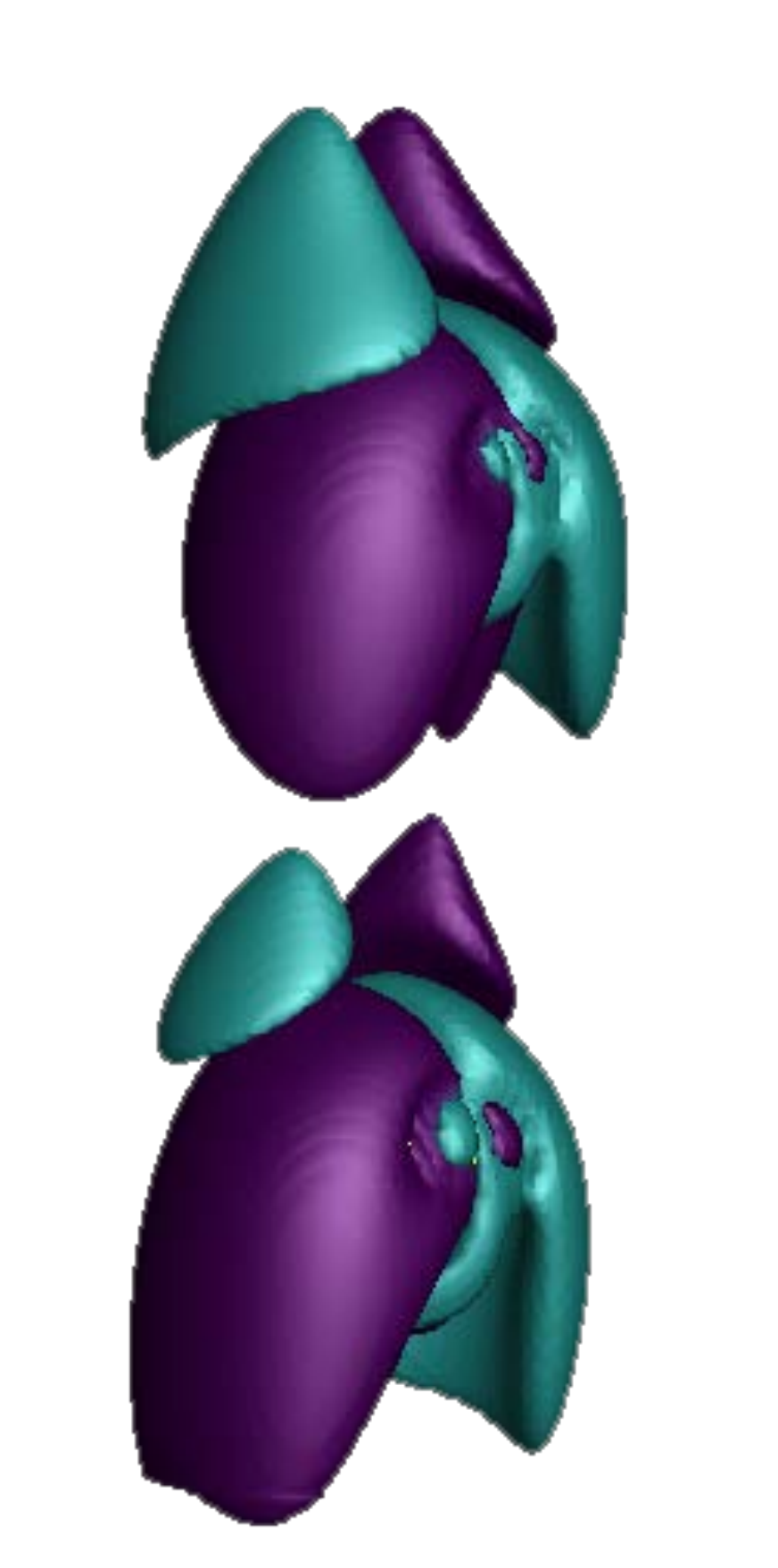}
}
\caption{Simulation of two interacting bubbles ascending in a fluid. (a) and (c) show the trajectories of the two bubbles, distances are in bubble radius units; (b) and (d) show streamwise iso-vorticity surfaces around the bubbles, $\omega_x U_B/R= \pm 0.05$. For both cases $Bo=0.05$, $Ar=400$; for (a-b) $La=0$ and for (b-c) $La=1.0$. See videos of these simulations in the Supplemental Material.}
\label{fig:simulations1} 
\end{figure*}

Fig. \ref{fig:simulations1} shows typical results for two characteristic cases. In Fig. \ref{fig:simulations1} (a) and (b) the results of a nearly spherical bubble pair ($Ar=400$) without surfactants ($La=0$) are shown. The bubbles start close to an in-line configuration at the start of the simulation; after some time, the bubbles no longer stay vertically aligned, moving away from each other as shown in Fig. \ref{fig:simulations1} (a). During this initial transit period, the wake behind the bubbles develops. To visualize the wake, we show  isovorticity surfaces of the stream-wise vorticity, as shown in Fig. \ref{fig:simulations1} (b). The interaction of the bubble with the wake in front of it leads to the appearance of a lift force as indicated by Eqn. \ref{eqn:lift}. For the case  of spherical bubbles and the absence of surfactants, the wake has two strands of counter-rotating vorticity, in agreement with \cite{adoua2009}. This wake structure induces a lift force in opposite directions for each subsequent bubble. Therefore, the bubbles are pushed sideways, leading to an increasing separation and deviation from the in-line configuration. These chains are unstable. 

In contrast, by considering also spherical bubbles ($Ar=400$) but including the effect of surfactants such that  $La=1$ the behaviour changes significantly. Fig.   \ref{fig:simulations1} (c) shows the trajectory for the two bubbles. Clearly, the initial in-line configuration remains unchanged even after the initial transient period. The structure of the wake around the bubbles in this case, shown in  Fig. \ref{fig:simulations1} (d), is significantly different from the case without surfactants. In particular, the vortex pair developing behind the trailing and leading bubble have the same rotation direction. This configuration induces a lift force of same direction for both bubbles; therefore, the lift force does not destabilize the in-line configuration.

As in the case of experiments, simulations were conducted for varying  $Ar$, $Bo$ and $La$ to identify the conditions that lead to stability or not.  In agreement with the experiments, stable chains were observed for larger bubbles (more deformed) or when surfactants were present. More importantly, since the entire flow field  around the bubble pair is know, the vorticity field could be readily obtained. As discussed below, the reversal of the lift force can be attributed to the amount of vorticity created at the bubble surface. Therefore, a criteria for stability can be proposed based on the maximum surface vorticity for each case. 

\section{Discussion}

For the range of Bond and Archimedes numbers relevant to bubbly drinks, the in-line clean bubble configuration is unstable. In other words, stable bubble chains cannot be observed because hydrodynamic interactions lead to dispersion. The dispersion is the result of the wake-induced lift force. Increasing the bubble injection frequency increases the lateral dispersion, as shown in Fig. \ref{fig:frequency}. An increase in the bubble frequency results in a smaller separation distance between successive bubbles, for which the wake-induced lift force is larger. Hallez and Legendre \cite{hallez_interaction_2011} showed that the magnitude of the wake-induced lift is proportional to the rotational part of the wake velocity field; the lift magnitude evolves as $1/(r \sin(\theta))$ where $r$ is the distance between the two bubbles centers and $\theta$ is the angle between the center line and the vertical. 

The experiments also showed that increasing the bubbles size, without surfactants, a stable bubble chain configuration could be observed, as shown in Fig. \ref{fig:exp1} (a). The bubble chain becomes stable because the lift force changes sign. The origin of the lift reversal {for deformed bubbles in a shear flow at large bubble Reynolds number} was explained by Adoua et al. \cite{adoua2009}. {The positive lift effect ($C_L > 0$ in Eq. \ref{eqn:lift}) results of the tilting and stretching mechanisms of the vorticity of the upstream  flow  in the bubble wake \cite{auton1987, legendre_lift_1998}. The vorticity produced at the bubble surface is also tilted and stretched in the wake; however, its contribution gives rise to a streamwise vorticity of opposite sign and thus an induced lift contribution of opposite direction. When the surface vorticity exceeds a certain amount, its contribution becomes dominant and therefore the  induced lift changes direction ($C_L<0$)}. As the bubble size increases, the bubble shape evolves becoming more oblate which, in turn, increases the production of surface vorticity. Hence, it is expected to observe a lift reversal as bubbles become more deformed. Such a change on the lift direction been observed experimentally {for  air bubbles in different fluids including water}\cite{hayashi_lift_2020, hayashi_lift_2021}. 

For the case of bubbles in champagne and other carbonated drinks the bubble size is too small to explain the {deformation-induced lift reversal. Typically, for a bubble in water, the lift reversal is observed for a bubble Reynolds number $Re \approx 1000$ corresponding to a bubble aspect ratio  $\chi\approx 2$  \cite{hayashi_lift_2021}}. The current experiments and simulations showed that a stable chains could be observed for small spherical bubbles if the surface was contaminated with surfactants. Therefore, we can argue that surfactants can also induce a reversal of the lift force. Such a reversal has been reported in previous experiments  \cite{takagi2008}. When a sufficient amount of surfactants is present, the bubble surface becomes immobile. In other words, the surface of the bubble can no longer sustain slip. Such non-slip condition leads to an increase of the production of vorticity at the surface which is transported to the bubble wake and is sufficient to change the wake structure. In this case, again, the lift force changes sign which implies that the trailing bubble in the wake is pushed to remain in-line forming a stable chain.

\begin{figure}[ht]
\centering
\subfigure[]{ \includegraphics[width=0.45\columnwidth]{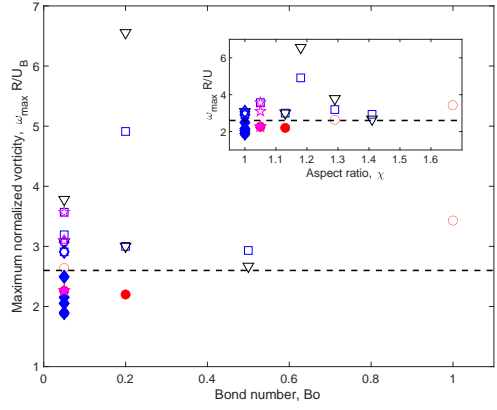}
}\\
\subfigure[]{ \includegraphics[width=0.45\columnwidth]{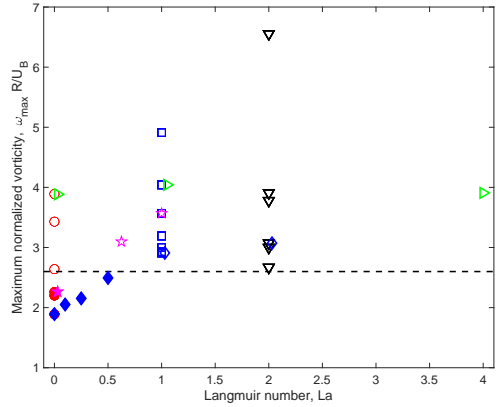}
}\caption{Maximum normalized azimuthal vorticity on bubble surface as function of Bond number (a) and Langmuir number (b). Filled and empty symbols show unstable and stable chains, respectively. The dashed line shows the threshold value of $\omega_\text{max}R/U_B$=2.65. The red circles, blue squares and black triangles shown simulations for different values of $Bo$ but $La=0$, $La=1$ or  $La=2$, respectively. The blue diamonds, magenta stars and green side-ways triangles show simulations for which both $Bo$ and $La$ were varied. Note that the $Ar$ was also varied in the simulations (400$<Ar<$11000).}
\label{fig:maxvorticity}
\end{figure}

Since we have access to all flow properties from the numerical simulations we can calculate the amount to vorticity created at the bubble surface.  As discussed above, we expect the surface vorticity to exceed a certain value for increasing bubble deformation, contamination or both.
 
{The deformation of an air bubble in water can be described in terms of the  Bond number, $Bo$, or the bubble aspect ratio, $\chi$ \cite{legendre2012deformation, hayashi_lift_2021}. } In Fig. \ref{fig:maxvorticity} (a) the maximum value of the surface vorticity for each simulation is shown as a function of the Bond number, $Bo$, and the bubble aspect ratio, $\chi$. For instance, for a certain bubble size, the Bond number is fixed. By progressively adding surfactants the maximum amount of surface vorticity increases. It is observed that for all cases the bubble pair becomes stable once maximum surface vorticity exceeds a threshold value of $\omega_\text{max}R/U_B$=2.65. This value can be surpassed by adding surfactants or by increasing $Bo$ or $\chi$. The plot also shows that either the aspect ratio or the bubble Bond number can be used to quantify the bubble deformation.

{Considering now the effect of surfactant at the interface, }in Fig. \ref{fig:maxvorticity} (b) the maximum vorticity is shown as a function of the Langmuir number for different values of the Bond number. As in the previous case, it is clear that the bubbles remain in an in-line configuration once  the surface vorticity exceeds the threshold value. Note that some additional simulations are shown here were the value of $La$ gradually increased for a fixed $Bo$ (blue diamonds). The results are in good agreement with the rest of the simulations.


As shown above, the threshold value for $\omega_\text{max} \sim 2.65 \,  U_B/R$ is observed at the transition from unstable to stable bubble line. The same vorticity amount is required to reverse the lift force when either bubble deformation or  bubble surface contamination are increased. This confirms that the transition from unstable to stable  is related to the reversal of the lift experienced  in the wake of bubble.

{To understand why this is the case, let us consider the stress balance at the bubble interface. At the bubble surface, the  viscous shear stress, \boldmath$\Sigma$, balances  the gradient of surface tension, $\nabla_I \sigma$:
\begin{equation}\label{Eq_cont1}
(\mathbf{\Sigma} \cdot \mathbf{n_b}) \times \mathbf{n_b} = \nabla_I \sigma,
\end{equation}
where $\mathbf{n_b}$ is the normal vector to the bubble surface.
The gradient of surface tension can be expressed as a function of the surfactant concentration using the Langmuir adsorption isotherm \cite{Levich1962}. The tangential viscous shear along the flow direction is  $ \mu \left({\partial u_I}/{\partial r} - { u_I}/{R_c} \right)$ and the interfacial azimuthal vorticity writes  $\omega_I  = {\partial u_I}/{\partial r} + { u_I}/{R_c}$ with  $u_I$ the local bubble interfacial velocity and $R_c$ the radius of curvature.  The interfacial vorticity can then be directly expressed as a function of the two above mentioned mechanisms, namely deformation and surface contamination, as
\begin{equation}\label{Eq_cont2}
\omega_I    =  2 \frac{ u_I}{R_c} - \frac{\mathcal{R} T \Gamma_\infty}{\mu \left(\Gamma_\infty- \Gamma \right)} \nabla_I \Gamma
\end{equation}
This relation shows that both the surface deformation and the surface contamination can increase the magnitude of the interfacial vorticity $\omega_I$. Indeed, deformation reduces at the bubble equator the radius of curvature $R_c$ and the surface contamination results in an accumulation of surfactants at the rear of the bubble.
This expression can be normalized by considering the bubble equivalent radius $R$, the characteristic rising velocity of a bubble $\rho g R^2/\mu$ and the  concentration of surfactant at equilibrium $\Gamma_0=\Gamma_\infty La/(1+La)$ (see section Material and Methods). The normalized vorticity $\omega_I^*$ (denoting all normalized variables with $^*$) at the interface can be written as:
\begin{equation}\label{Eq_cont23}
\omega_I^*    =  2 \frac{ u_I^*}{R_c^*} - \frac{La \Lambda }{1 + La (1-  \Gamma^*)} \nabla_I^* \Gamma^*.
\end{equation}
}

With these ideas, we can now propose a stability map that considers both the effect of bubble deformation and surface mobility {(first and second terms in the right hand side of Eq. \ref{Eq_cont23}, respectively)} as the main factors determining the stability of a bubble chain. The bubble deformation can be readily quantified by the $Bo$ number. While $La$ quantifies the amount of surfactants {that can be transferred from} the liquid to the bubble surface, {a better measure of the effect of surfactants on the interface behavior is given by the parameter $La \Lambda $, as demonstrated by Eq. \ref{Eq_cont23}. }

\begin{figure}[ht]
\centering
\includegraphics[width=0.9\columnwidth]{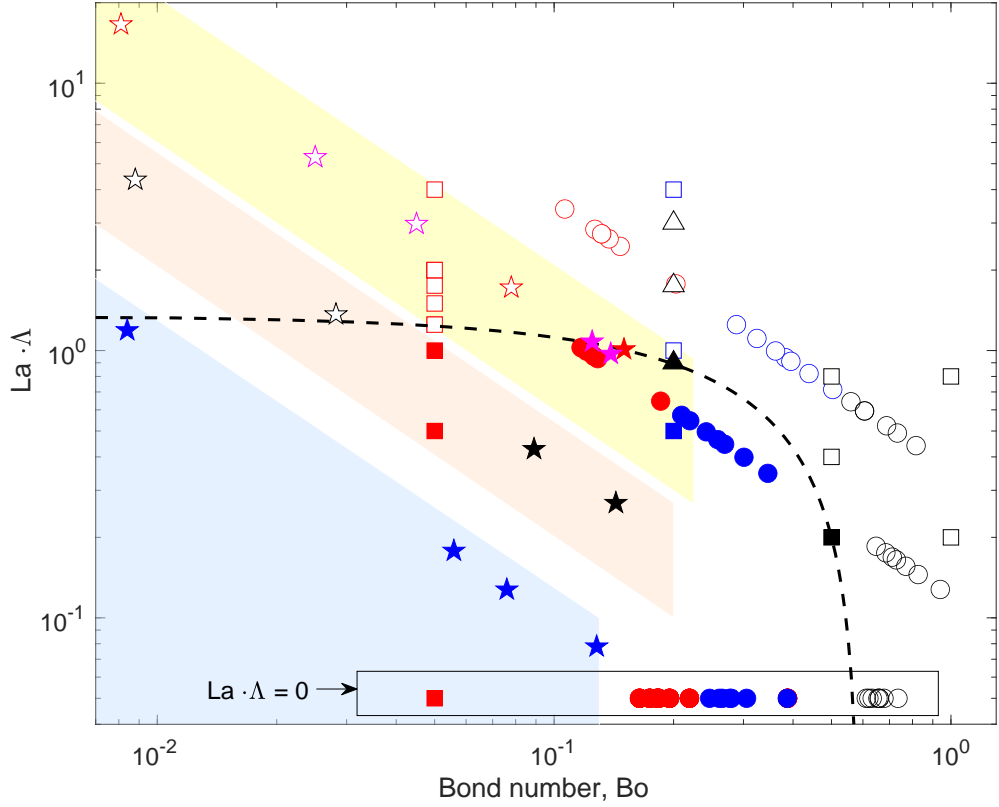}
\caption{Stability map showing the stable and unstable chain behaviour in terms of the Bond number and the {product of the rigidity parameter with the Langmuir number}. Filled and empty symbols show unstable and stable chains, respectively. Circles and stars show experimental data, while squares and triangles show numerical results. The data enclosed in the rectangle corresponds to cases for $La\Lambda=0$ (clean fluids). The dashed line presents an line for which $\omega_\text{max} R/U_B\approx2.65$ which marks the transition. The stars (blue, black, red and magenta) show the measurements conducted with carbonated water, beer, sparkling wine and champagne, respectively. The yellow, orange and blue shaded areas represent the expected range of values for champagne, beer and water, respectively, according to Table \ref{tab:table1}.}
\label{fig:stabilitymap}
\end{figure}

Fig. \ref{fig:stabilitymap} shows both experiments and simulation results in a map of $Bo$ and $La \Lambda$. The filled and empty markers show unstable and stable chains, respectively. The combination of bubble deformation and surface rigidity clearly determine the stability conditions of a bubble chain. Furthermore, the transition results from the amount of vorticity created at the bubble surface as indicated by the threshold value (dashed line). If the contamination level and the bubble size are known, the map presented here can be used to determine the stability of a bubble chain. Again, small bubbles, as those typically observed in carbonated drinks are not expected to be stable in clean liquids. However, the presence of surfactants leads to surface immobilization to result in stable chains. 

We can also note that when bubbles are very small, for $Ar<1$, a wake will not develop \cite{clift2005bubbles} and therefore, regardless of contamination, a lift force does not appear. In aqueous liquids, an $Ar<1$ corresponds to $Bo<10^{-3}$. Bubble chains to the left of this value in the map shown in Fig. \ref{fig:stabilitymap} will be stable.

In this study, we neglect the effect of bubble growth. Since the growth rate of the bubbles is smaller than the bubbles velocity, the flow field around the bubbles is decoupled from the bubble growth, i.e, at every moment, the flow field is at equilibrium with the bubble size and shape. Therefore, Fig. \ref{fig:stabilitymap} can be used to predict the evolution of chain stability as the bubbles grow. Indeed, in that situation, $Bo$ and $Ar$ increase as the bubble rise; however, $La \Lambda$ decreases as the bubble grows. Therefore, the chain may transition from stable to unstable and vice versa, as the bubbles grow.

To end and to sustain that these claims are valid for bubbles in typical carbonated drinks, including champagne, we attempt to quantify the Bond number, the rigidity parameters and the Langmuir number for three types of carbonated beverages. We also conduct experiments with actual beverages. The calculation of $Bo$ is straight forward, since the values of density, size and surface tension are known. For the calculation of $\Lambda$ and $La$, on the other hand, some assumptions must be made of the surface absorption and nature of the typical surfactants in these beverages (see Methods section for details). Considering bubbles ranging in radius from 0.25 to 1 mm, carbonated water is not expected to produce stable chains since a small amount of surfactants is present. The blue shaded area in Fig. \ref{fig:stabilitymap} show the range of parameters valid for water. For champagne, the type of surfactants and the quantities are well studied and known. Hence, it is possible to confidently assign the range of expected values of $\Lambda$ and $Bo$ for the same range of sizes. The yellow shaded region, shown in Fig. \ref{fig:stabilitymap}, clearly indicates that for this beverage stability of bubble chains is expected for small bubbles. This is in agreement with what is typically observed. For the case of beer, on the other hand, we had to make assumptions to infer $\Lambda$, since the typical surfactants are very different. We predict that for the same range of bubbles sizes, denoted by the orange-shaded area in the figure, bubble chains can either be stable or not. This observation can be seen in some beer glasses, but more carefully conducted experiments are needed to assess bubble chain stability and to understand how proteins immobilize the surface of bubbles in beer. 

The experiments conducted with very small bubbles in water, beer, sparkling wine and champagne are also shown in Fig. \ref{fig:stabilitymap}, depicted by the stars. Special glass-pulled capillaries were fabricated to produce bubbles in the size range of actual beverages. The sizes and values of the relevant parameters are shown in Table \ref{tab:table_props}. A video is also available in the Supplemental Materials. The chains were stable for sparkling wine and champagne, considering small bubbles. Note that in one case, we were able to produce a weakly unstable bubble chain for sparkling wine  ($Bo=0.13$). For beer, the smallest bubble size produced stable chains, while the other two sizes showed unstable chains. For water, experiments with these small bubble sizes were all unstable. These observation are in good agreement with our experiments with mock fluids and numerical simulations.

\section{Conclusions}
In summary, we have demonstrated that the stable bubble chains typically observed in champagne occur due to the presence of surfactants. For certain bubble sizes, the bubbles can only move in-line if enough vorticity is produced on their surface. If bubbles are small, the deformation is not enough to produce sufficient vorticity. Hence, surface immobilization (resulting from the presence of surfactants) provides a second source of vorticity that leads to a stable chain. The current results explain why one observes stable bubble chains in champagne, but also provide an interesting case study of sources of vorticity in bubbly two-phase flows. Additionally, since these effects are now identified, the stability of a bubble chain could be used to assess the level of surfactant contamination in any liquid. This idea could be of practical use.

\begin{acknowledgments}
The computational time was furnished by the CALMIP scientific group (project number P17006), whose contribution is greatly appreciated. O.A thanks Jan van Nieuwkasteele and the "Soft Matter Fluidics and Interfaces" group of University of Twente for the computational resources used for development and postprocessing purposes.
\end{acknowledgments}

\bibliographystyle{unsrt}
\bibliography{champagne_references}

\end{document}